\documentclass[12pt,draftclsnofoot,a4paper,onecolumn]{ieeetran}

\usepackage[utf8]{inputenc} 

\def\x{\mathbf{x}}
\def\y{\mathbf{y}}

\def\H{\mathbf{H}}

\def\n{\mathbf{n}}

\def\v{\mathbf{v}}
\def\u{\mathbf{u}}
\def\C{{\mathbb C}}

\usepackage{amsmath,amsfonts,amssymb}

\usepackage{graphicx}
\graphicspath{{./figures/}}

\usepackage{url}
\usepackage{algorithmic}
\usepackage{algorithm}
\usepackage{makecell}
\usepackage{booktabs}

\usepackage{xcolor}

\newcommand{\myfigurewidth}{0.75\textwidth}

\begin{document}

\title{Experimental Evaluation of Interference Alignment for Broadband WLAN Systems}

\author{Christian~Lameiro,~\IEEEmembership{Student Member,~IEEE,}
        \'Oscar~Gonz\'alez,~\IEEEmembership{Student Member,~IEEE,}
 		Jos\'e~A.~Garc\'ia-Naya~\IEEEmembership{Member,~IEEE}
        Ignacio~Santamar\'ia,~\IEEEmembership{Senior~Member,~IEEE},
        and~Luis~Castedo~\IEEEmembership{Member,~IEEE}
\thanks{C.~Lameiro, \'O.~Gonz\'alez and I.~Santamar\'ia are with the Department of Communications
Engineering, University of Cantabria, 39005 Santander, Spain
(e-mail: \{lameiro, oscargf, nacho\}@gtas.dicom.unican.es).}
\thanks{J.A.~Garc\'ia-Naya and L.~Castedo are with the Department of Electronics and Systems, University of A Coru\~na, 15071 A Coru\~na, Spain
(e-mail: \{jagarcia, luis\}@udc.es).}}

\maketitle

\begin{abstract} 
{In this paper we present an experimental study on the performance of spatial Interference Alignment (IA) in indoor wireless local area network scenarios that use Orthogonal Frequency Division Multiplexing (OFDM) according to the physical-layer specifications of the IEEE 802.11a standard. Experiments have been carried out using a wireless network testbed capable of  implementing a 3-user MIMO interference channel. We have implemented IA decoding schemes that can be designed according to distinct criteria (e.g. zero-forcing or MaxSINR). The measurement methodology has been validated considering practical issues like the number of OFDM training symbols used for channel estimation or feedback time. In case of asynchronous users, a time-domain IA decoding filter is also compared to its frequency-domain counterpart. We also evaluated the performance of IA from bit error rate measurement-based results in comparison to different time-division multiple access transmission schemes. The comparison includes single- and multiple-antenna systems transmitting over the dominant mode of the MIMO channel. Our results indicate that spatial IA is suitable for practical indoor scenarios in which wireless channels often exhibit relatively large coherence times.}
\end{abstract}


\section{Introduction} \label{sec:Introduction}
Interference management is a key issue in the design of wireless systems. When several users transmit over the same wireless resources, orthogonal access techniques such as Frequency-Division or Time-Division Multiple Access (FDMA and TDMA, respectively) are traditionally applied to avoid interference among them. In orthogonal multiple access schemes the system bandwidth and/or time resources are divided among users and the individual data rates decrease with the network size. Interference Alignment (IA) has been recently proposed as an alternative interference management method that confines interference signals within half of the signal space at each receiver, hence allowing each user to transmit over the interference-free subspace~\cite{Jafar2008}.

Although a large number of theoretical results have shown IA to be a very promising technique, there is still lack of experimental Over-The-Air (OTA) results evaluating its actual performance in real wireless scenarios. This scarcity of experimental results is mainly due to the high costs and effort required to conduct OTA measurements in IA scenarios. For example, to evaluate the practical performance of spatial IA methods, at least six nodes (three transmitters and three receivers) with two antennas each are needed to implement the simplest Multiple-Input Multiple-Output (MIMO) interference channel.

\subsection{Previous Experimental Work on IA}
The first work that tackled a real-world implementation of IA was presented in \cite{IAC}. This work considered the implementation of IA techniques combined with cancellation methods over a wireless network testbed comprised of 20 Universal Software Radio Peripheral (USRP) nodes equipped with two antennas each. The implemented technique does not correspond to pure IA because it requires a certain amount of cooperation among access points in such a way that all the network interference can be nulled out. Several practical issues were addressed in this work, showing that IA is unaffected by frequency offsets or by the use of different modulations. Imperfect time synchronization, however, affects IA but this issue can be overcome by performing IA at the sample level, i.e., before demodulation and synchronization takes place. Finally, this work posed the interesting question of how to perform sample level alignment in Orthogonal Frequency-Division Multiplexing (OFDM) systems over frequency-selective channels.

IA was further evaluated in \cite{OmarTVT2010}, where the authors conducted an experimental study over measured indoor and outdoor MIMO-OFDM channels. By modifying the distance among network nodes and antennas, they characterized the effect of spatial correlation and subspace distance, and showed that IA is able to achieve the maximum available Degrees of Freedom (DoF) over realistic channels. However, although the channels were obtained from measurements, no OTA transmissions of aligned signals were actually measured. Therefore, many practical issues such as time/frequency synchronization, imperfect Channel State Information (CSI), and dirty Radio-Frequency (RF) effects such as phase noise, non-linearities, IQ imbalance, or clipping and quantization in the Digital-to-Analog and Analog-to-Digital converter (DAC/ADC) were not taken into account. In \cite{IWSSIP2012}, different IA schemes are evaluated in the 3-user Single-Input Single-Output (SISO) interference channel using frequency extensions. As in \cite{OmarTVT2010}, the results in \cite{IWSSIP2012} were obtained using urban macro-cell measured channels but without transmitting aligned signals.

In~\cite{WSA_2011,Eusipco2011_IA,Zetterberg2011} the first aligned real transmissions were conducted to evaluate spatial-domain IA in a 3-user interference channel, thus providing more precise results about the actual performance of IA in realistic scenarios. In~\cite{WSA_2011}, the feasibility of spatial IA over indoor channels and single-carrier transmissions was studied, identifying also some practical issues that affect IA performance already pointed out in \cite{IAC} and \cite{OmarTVT2010}. In addition, the CSI estimation error was also described as an important issue, that was further analyzed in \cite{Eusipco2011_IA}. The 3-user MIMO interference channel with OFDM transmissions is also studied in \cite{Zetterberg2011}, along with coordinated multi-point transmission methods. In this work RF impairments are identified as an important source of mismatch between practical and theoretical performance of IA. However, the work in \cite{Zetterberg2011} focuses on verifying simulation models and no analysis of the inherent limitations of IA is performed.

The work in \cite{massey2012} described two real-time implementations of IA in a 3-user MIMO-OFDM scenario showing that the computational power of current embedded platforms makes software-defined implementations of IA feasible. Another approach was followed by the authors of \cite{balan2012}, where blind IA was implemented with the aim of avoiding the intense global CSI requirements of spatial-domain IA.

{Recent experimental evaluations of spatial interference alignment analyze the main performance limiting factors found in real-world scenarios \cite{SAM2014_Mayer}, study the impact of outdated channel state information \cite{SAM2014_Artner}, consider its combination with antenna selection techniques \cite{EW2014_ElAbsi}, or consider analog feedback in a distributed real-time implementation \cite{TVT2014_Lee}}.

\subsection{Summary of Key Practical Issues Arising when Implementing IA Techniques}
Despite the promising theoretical results on IA, several practical impairments come up in real scenarios that might degrade the overall system performance. In the following we detail the main issues affecting practical IA transmissions.

\subsubsection{Imperfect CSI}

IA is usually studied assuming perfect CSI is available at every node of the network, a premise that never occurs in practice. Moreover, since the computation of the precoders and decoders involves all the pairwise interference channels, even a slight time variation of a single channel would ideally result in a change of all IA precoders and decoders. In practice, this causes two problems. First, the presence of channel estimation errors or time variations makes it impossible to perfectly suppress interference~\cite{WSA_2011,Eusipco2011_IA}. Second, nodes must exchange its local CSI to compute the IA solution, and this introduces additional overhead and delay between channel estimation and data transmission. During this elapsed time, the channel may vary, hence out-dating the CSI estimates especially when there are moving scatterers in the surroundings. Besides CSI estimation errors, dirty RF effects~\cite{FetweissDirtyRF2005} are also responsible for a great portion of the gap between ideal and practical setups. Major contributors to distortion in OFDM systems are nonlinear amplifiers, clipping, ADC effects and phase-noise. Some of these effects have been modeled in~\cite{Zetterberg2011}.

\subsubsection{Signal Collinearity}

Even under the unrealistic assumption that perfect CSI is available, the received signal is projected into the subspace orthogonal to the interference in order to null the interferences once they are aligned. In this process, part of the desired signal energy is lost due to spatial collinearity between signal and interference subspaces. In the presence of high spatial collinearity, the desired signal power is severely reduced. To overcome this problem, many algorithms have been proposed to reach a trade-off between signal and interference power, such as the MaxSINR algorithm \cite{AltMinIA} considered in this work. Recent works have also suggested the use of antenna switching strategies \cite{EW2014_ElAbsi,elHadidy2012,Bahl2012}.

\subsubsection{Synchronization}\label{sec:synchronizationIssues}

We consider the following scenario: OFDM spatial interference alignment transmissions in the 3-user MIMO interference channel with two antennas per transmit/receive node; a single data stream per user; a wireless local area network in indoor environments; and the IEEE 802.11a waveform –-with a bandwidth of 20\,MHz-– for the three users. Assuming that the same waveform (including training and/or preamble sequences) is employed at the three transmitters, we then distinguish two different cases with respect to synchronization:
\begin{itemize}
\item[I)] All users are perfectly synchronized in time and frequency. This means that the three transmitters transmit exactly at the same time instants, while the three receivers are able to perfectly acquire the time and frequency references of its corresponding transmitter before processing the received signals. Another possibility would consist in assigning orthogonal training sequences to all users. Given that the transmitters operate synchronously, each receiver acquires the time reference with respect to its desired transmitter without being affected by interferences.

\item[II)] All users operate asynchronously in an uncoordinated way, leading to symbol timing offsets between the desired and the interfering OFDM symbols. Consequently, each receiver has to acquire the time and frequency reference from the receive signal, which consists of the desired signal plus the interference from two of the three transmitters. Therefore, the Signal-to-Interference Noise Ratio (SINR) at the input of the receiver decreases with respect to case I. Notice that assigning orthogonal training sequences to the users does not alleviate the problem because now the transmission of those orthogonal sequences is not synchronous. Consequently, it cannot be guaranteed that the observed training sequence at the receiver is not affected by interferences from the other users.
\end{itemize}

With respect to spatial interference alignment, precoding at the transmitter and decoding at the receiver can be performed in the frequency domain (the usual approach in the literature) or in the time domain, leading to a set of four different possibilities to apply spatial interference alignment precoding and decoding.
Notice that, as shown in \cite{ISWCS2012}, interference leakage at the OFDM receiver is completely independent of the delays between the transmitters and the receivers only if spatial interference alignment precoding and decoding operations are carried out in the time domain. Otherwise, the magnitude of the interference leakage will depend on to the delays between transmitters and receivers. Even if those delays are apparently small (e.g., 2 samples at 40\,MHz sampling frequency yielding 50 nanoseconds due to an imperfect synchronization among the three transmitters), they may produce interference leakage after decoding at the receiver. {The imperfect cancelation occurs when some samples of the undesired users adjacent OFDM symbols interfere the current one due to an insufficient Cyclic Prefix (CP) length or time misalignments. This is because the frequency-domain IA scheme is designed to cancel the interference when the system can be equivalently decomposed into a set of non-overlapping channels. If this is not the case, there will be Inter-Symbol and Inter-Carrier Interference (ISI and ICI, respectively) components in the interfering signals that cannot be eliminated.}

Spatial interference alignment decoding in the time domain consists in filtering the received signal in the time domain. On the other hand, time-domain decoding cannot effectively suppress all the interference because of the resulting filter length. The length of this filter can easily introduce inter-symbol interference as it may exceed the CP length minus the channel delay spread \cite{ISWCS2012}.

Therefore, interference can be completely suppressed at the receiver only when interference alignment precoders and decoders are applied at the frequency domain on a per-subcarrier basis, hence demanding for a synchronous scenario.

Contrarily, if a fully synchronous scenario is not feasible, the SINR at the input of the receiver decreases due to the high level of interference. On the other hand, it is well known that the performance of synchronization tasks depends on the SINR at the receiver input, and therefore their performance will improve if the SINR is increased by reducing the level of interference. This can be achieved by applying interference alignment decoding in the time domain at the receiver input, before the synchronization tasks. However, there is a trade-off between the level of inter-symbol interference introduced by the time-domain filtering and the multiuser interference-suppression capacity. The longer the interference-suppression filter in the time domain, the higher the inter-symbol interference and the lower the interference leakage.

In the light of the above-mentioned comments, one could think that interference alignment in the time domain is much more convenient than in the frequency domain in totally asynchronous scenarios, as interference leakage due to delays between transmitters and receivers can be completely avoided and, at the same time, synchronization tasks perform at higher SINR levels. However, for the 3-user scenario under consideration, closed-form and computationally efficient solutions do exist for spatial interference alignment in the frequency domain, but not in the time domain \cite{Jafar2008}. The solution presented in \cite{ISWCS2012} for calculating optimum time-domain interference alignment precoders and decoders is computationally expensive, leading to much longer (and with larger variance) feedback times. A way to alleviate the problem consists in computing the IFFT of the frequency-domain solutions and truncating the resulting filters to achieve a good trade-off between interference suppression and ISI.

\subsection{Contributions}\label{sec:contributions}
In this paper, we extend our work in~\cite{WSA_2011,Eusipco2011_IA} to broadband OFDM wireless transmissions. Specifically, we use the IEEE 802.11a Wireless Local Area Network (WLAN) physical-layer standard~\cite{80211a} as a benchmark to evaluate the performance of spatial IA in an illustrative indoor scenario. The measurement setup can be thought of as an indoor WLAN system in which three access points ---with two antennas each--- communicate simultaneously over the same frequency (channel resource) with three static devices (e.g.\ laptops), also equipped with two antennas each. This is opposed to conventional WLAN systems that would assign different channels to each communication link (i.e., FDMA). The goal of this paper is to evaluate experimentally several spatial IA schemes in indoor WLAN applications, identifying and analyzing the main issues that degrade their performance in broadband OFDM transmissions, and comparing their end-to-end measurements with those of TDMA-based schemes. In this work, we only consider systems in which each user transmits a single stream of data to its intended receiver. More specifically, the main contributions of this work are the following:
\begin{itemize}
\item With respect to our previous work in \cite{WSA_2011} and \cite{Eusipco2011_IA}, in which only single-carrier transmissions over flat-fading channels were considered, here the experimental work focuses on OFDM transmissions based on the 802.11a standard and with a 20\,MHz bandwidth. Broadband transmissions pose new difficulties but also permit the implementation of more complex IA schemes. Additionally, we have improved our measurement methodology and we have also reduced the time elapsed between channel estimation and IA transmission from 5 seconds in \cite{WSA_2011} and \cite{Eusipco2011_IA} to a second.

\item As discussed previously, we consider and compare the performance of spatial IA decoding schemes that operate either in time domain\cite{ISWCS2012,ICASSP2012}, or in a more conventional per-subcarrier basis in frequency domain. Furthermore, we have assessed the actual performance of IA and MaxSINR schemes \cite{AltMinIA}.

\item Additionally, we analyze the main issues that might affect our measurement methodology (see Section~\ref{sec:methodology}), and consequently our results, such as the number of training symbols used for channel estimation or the feedback time elapsed between training and transmissions of aligned frames.

\item Finally, we present Error Vector Magnitude (EVM) and Bit Error Rate (BER) measurement-based results for different data rates. We also compare them to those obtained when TDMA-based transmissions are employed. The comparison includes SISO and MIMO systems transmitting over the dominant mode of the MIMO channel (referred to as Dominant Eigenmode Transmission or DET \cite{Andersen}).

\end{itemize}

Taking into account all pros and cons outlined in Section~\ref{sec:synchronizationIssues}, we always apply the spatial interference alignment precoders at the transmitters in the frequency domain on a per-subcarrier basis, while the three transmitters and receivers are synchronized between them in time (up to 2 samples at 40 MHz sampling frequency) and in frequency due to the following reasons:
\begin{enumerate}
\item to keep the feedback time short (and with low variability) during the measurements;
\item to employ the same transmit waveform for the three users (i.e., training signals do not depend on the number of users);
\item to reuse (without a significant performance degradation) conventional time and frequency synchronization algorithms valid for OFDM-based wireless systems in single-user interference-free scenarios;
\item to be able to compare the performance of spatial interference alignment decoding at the receiver applied in time with respect to when it is applied in frequency under the following conditions:
	\begin{enumerate}
	\item the same set of spatial interference alignment precoder and decoder vectors (i.e., the same interference alignment solution) is employed;
	\item the aforementioned interference alignment solution was computed from the same channel realization;
	\item the same set of acquired frames experiencing the same channel realizations is used to estimate the considered figures of merit (EVM, BER) when interference alignment decoding is applied in time domain or in frequency domain;
	\item time synchronization is performed when interference alignment is applied in time domain and reused for the frequency-domain case, hence the performance of the frequency-domain interference alignment decoding is not degraded because of the interference.
	\end{enumerate}
\end{enumerate}

We have carefully designed a measurement methodology (see Section~\ref{sec:methodology}) to be able to assess the performance of spatial interference alignment both in time and frequency domain. For comparison purposes, we have also evaluated the performance offered by other approaches like MaxSINR, SISO-TDMA, and DET-TDMA. Such a methodology also allows us to measure the interference leakage at the receiver, as well as to evaluate the performance of the aforementioned methods in the absence of interference. We also ensure that our measurement scenario is suitable for the application of the spatial interference alignment problem (high SNR at the receiver and interference levels compared to those of the desired signals, see Fig.~\ref{fig:fdp_SNR}). Finally, the validity of the above-mentioned measurement methodology is also analyzed to ensure that our comparison is not affected by insufficient training (see Fig.~\ref{fig:EVM_vs_Npilots}) or by excessive feedback time (see Fig.~\ref{fig:EVM_vs_feedbackTime}).

The rest of the paper is organized as follows. Section~\ref{sec:IA} describes spatial IA in a 3-user $2\times2$ MIMO-OFDM channel considering both post-FFT and pre-FFT IA decoding schemes. In Section~\ref{sec:testbed}, the wireless network testbed utilized for the measurements is briefly described. Measurement set-up and methodology are both explained in Section~\ref{sec:setup} and Section~\ref{sec:methodology}, respectively. The obtained results are discussed in Section~\ref{sec:results}. Finally, Section~\ref{sec:conclusions} concludes the paper.

\begin{figure}[!t]
\centering
\includegraphics[width=\myfigurewidth]{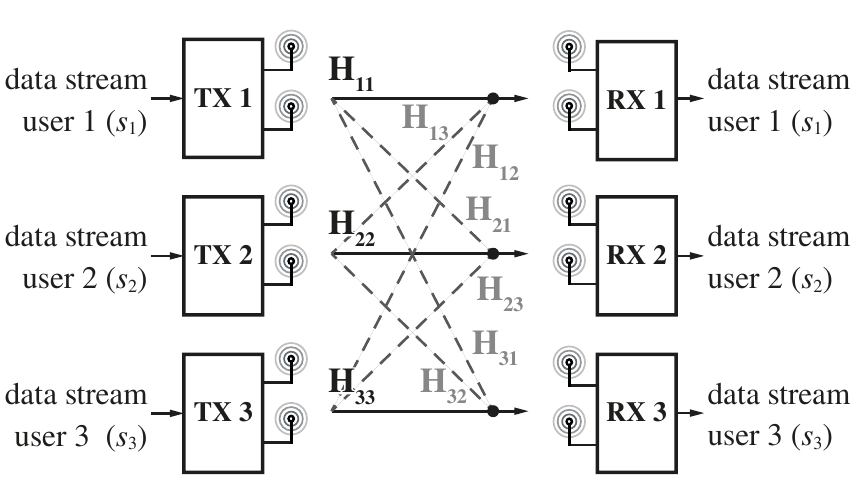}
\caption{{Scheme of the $(2\times2, 1)^3$ interference network. } Direct channel links are shown in black solid lines whereas interference channels are plotted in gray dashed lines.}
\label{fig:IAScheme}
\end{figure}

\section{Spatial Interference Alignment}\label{sec:IA}
IA is able to exploit the multiple time, frequency and spatial dimensions available in a wireless system. However, when aligning over the frequency or time domain, the number of required dimensions to arbitrarily approach to the maximum DoF promised by IA grows exponentially with the number of users \cite{JafarTut}. The number of required dimensions, on the contrary, is considerably less when aligning interference over the spatial dimension \cite{Gonzalez2014,Razaviyayn2012}; which facilitates its practical implementation. For instance, for a 3-user channel, $3n+1$ symbols can be transmitted using $2n+1$ extensions, where $n$ is an integer, yielding a total of $(3n+1)/(2n+1)$ DoF \cite{Jafar2008}. This would require a theoretically infinite number of frequency domain extensions to achieve the maximum number of 3/2 DoF in the 3-user SISO interference channel, while spatial domain IA is able to achieve the maximum number of 3 DoF with constant channels and two antennas. Furthermore, IA by means of symbol extensions requires significant multipath \cite{IAfeed,IAfeas}, whereas a sufficient antenna separation ensures no DoF loss when IA is performed in the spatial domain. Another advantage of spatial IA is that it can be readily applied while being compliant with any OFDM signaling format such as the 802.11a WLAN standard, as shown in this paper. On the contrary, any alignment scheme over time or frequency would require major changes on the physical layer format. Further, we focus on the $2 \times 2$ MIMO 3-user interference channel because it can be easily implemented with the multiuser MIMO testbed described in Section~\ref{sec:testbed}.

This section reviews the concept of IA in the spatial domain and discusses the application and design of IA decoders in the time and in the frequency domain. However, as commented in Section~\ref{sec:contributions}, in all experimental evaluations the IA precoders were always applied at the transmitters before the IFFT on a per-subcarrier basis in the frequency domain, whereas at the receivers the decoders are applied either in the time domain (pre-FFT decoding) or in the frequency domain (post-FFT decoding).

\begin{figure*}[t!]
\centering
\includegraphics{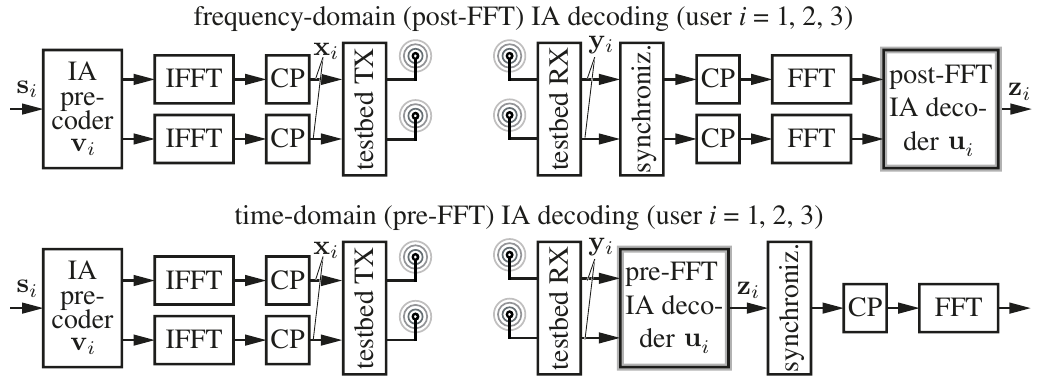}
\caption{{Top: frequency-domain (Post-FFT) IA decoding. Bottom: time-domain (pre-FFT) IA decoding.} Schematic of an OFDM transmitter that applies IA precoding in the frequency domain on a per-subcarrier basis. IA decoding is carried out in the frequency domain (top) and denoted as post-FFT IA decoding; or in the time domain (bottom), denoted as pre-FFT IA decoding.}
\label{fig:SchematicPostPreFFT}
\end{figure*}

\subsection{Interference Alignment with Post-FFT Decoding}\label{subsec:IA_postFFT}
Let us consider a 3-user MIMO interference channel comprised of three transmitter-receiver pairs (links) that interfere with each other as shown in Fig.~\ref{fig:IAScheme}. Each user is equipped with two antennas at both sides of the link and sends a single stream of data. Following the convention introduced in~\cite{Yetis2010}, this interference network is denoted as $(2\times 2,1)^3$. Assuming a fully coordinated scenario in which all users transmit their OFDM symbols exactly at the same time instants, or when the possible delays among users can be accommodated by the CP minus the channel delay spread, each receiver can use a conventional synchronizer and, consequently, the IA decoder can be applied after the FFT block on a carrier-by-carrier basis (see top of Fig.~\ref{fig:SchematicPostPreFFT}).

Assuming that the CP is sufficiently long to accommodate the channel delay spread, the discrete-time signal $\y_i$ at the $i$-th receiver for a given subcarrier (to not overload the notation unnecessarily, the index denoting the subcarrier is omitted in this section) is the superposition of the signals transmitted by the three users, weighted by their respective channel matrices and affected by noise, i.e.,
\begin{equation}
    \y_i=\H_{ii} \x_i + \sum_{j\neq i} \H_{ij} \x_j + \n_i,
\label{Eq:received_signal_narrowband}
\end{equation}
where $\x_i \in \mathbb{C}^{2 \times 1}$ is the signal transmitted by the $i$-th user, $\H_{ij}$ is the $2 \times 2$ flat-fading MIMO channel from transmitter $j$ to receiver $i$, and $\n_i \in \mathbb{C}^{2 \times 1}$ is the additive noise at receiver $i$.

\subsubsection{Closed-form Interference Alignment Solution}\label{sec:ClosedFormIA}
Spatial IA uses a set of beamforming vectors (precoders) $\{ \v_i \in \mathbb{C}^{2 \times 1}\}$ and interference-suppression vectors (decoders) $\{ \u_i \in \mathbb{C}^{2 \times 1}\}$ that must satisfy the following zero-forcing conditions for all transmitter-receiver pairs $i=1, 2, 3$:
\begin{equation}
\begin{cases}
\u_i^H \H_{ii} \v_{i} \neq 0 & \\
\u_i^H\H_{ij} \v_{j}=0,&\forall j\neq i.
\end{cases}
\label{eq:perfect_aligment_conditions}
\end{equation}
There is an analytical procedure to obtain precoders and decoders for the $(2 \times 2,1)^3$ case \cite{Jafar2008}:
\begin{enumerate}
   \item The precoder for user 1, ${\v}_{1}$, is any eigenvector of the following $2 \times 2$ matrix $\mathbf{E}$ (each eigenvector yields a different IA solution):
   \begin{equation}
   \label{eq:matrix_E}
     \mathbf{E}=(\H_{31})^{-1} \H_{32} (\H_{12})^{-1}\H_{13} (\H_{23})^{-1}{\H}_{21}.
   \end{equation}

   \item The precoders for users 2 and 3, ${\v}_{2}$ and ${\v}_{3}$, are respectively obtained as
   \begin{equation}\label{eq:v2}
     {\v}_{2} = ({\H}_{32})^{-1}{\H}_{31}{\v}_{1},\ \text{and}
   \end{equation}
   \begin{equation}\label{eq:v3}
     {\v}_{3} = ({\H}_{23})^{-1}{\H}_{21}{\v}_{1}.
   \end{equation}

   Since $\mathbf{E}$ is a full-rank $2 \times 2$ matrix with probability one for generic MIMO channels, in which each entry of the channel matrix is an independent and identically distributed random variable drawn from a continuous distribution, it has two eigenvectors that can be chosen as the precoder for the first user, hence yielding two distinct IA solutions. An interesting fact of the $3$-user interference channel is that it induces a permutation structure and, consequently, the procedure described above leads to exactly the same set of IA solutions regardless of the user employed for starting the procedure. In summary, there are only two different IA solutions per subcarrier.

    \item Finally, the interference-suppression filters (decoders) are designed to lie in the orthogonal subspace of the received interference signal, i.e., the decoder of user 1 is the eigenvector of $[\H_{12}\v_2 \; , \; \H_{13}\v_3]$ associated with the zero eigenvalue. The decoders for users 2 and 3 are obtained in an analogous way.
 \end{enumerate}
When zero-forcing IA linear precoders and decoders are applied at both sides of the link, the signal received by the $i$-th user is given by
  \begin{align}\label{eq:rx_signal_proj}
      z_i&=\u_i^H \H_{ii} \v_i s_i + \sum_{j\neq i} \u_i^H \H_{ij} \v_j s_j+ \u_i^H \n_i \notag \\
      &=\u_i^H \H_{ii} \v_i s_i + \u_i^H \n_i,
  \end{align}
where $s_i$ is the transmitted symbol corresponding to the $i$-th user. Notice that the signal from the $i$-th transmitter to the $i$-th receiver travels over the equivalent SISO channel $\u_i^H \H_{ii} \v_i$. The interference terms are totally suppressed when projecting the received signal onto the subspace whose basis is $\u_i$.

Similarly to zero-forcing channel equalization, zero-forcing IA suffers from noise amplification when MIMO channels are close to singular. Other approaches can be used to mitigate this limitation and perform better in the medium and low Signal-to-Noise Ratio (SNR) regimes. One such example is the MaxSINR algorithm \cite{AltMinIA} which has also been adopted in the measurements of this work for comparison purposes.

\subsubsection{MaxSINR Algorithm}\label{sec:MaxSINR}
{The MaxSINR algorithm aims at maximizing the SINR at each receiver by a proper design of the precoding and decoding vectors. As a result, it usually outperforms pure IA for medium and low SNR values, whereas it approaches IA as the SNR increases. To compute such precoders and decoders, an alternating optimization algorithm must be applied. Thereby the decoders (precoders) are optimized at each iteration while the precoders (decoders) are kept fixed. This results in a closed-form solution at each step of the algorithm. The MaxSINR algorithm can then be summarized as follows.

\begin{enumerate}
    \item While the precoders are kept fixed, choose the decoder of each user as the one that maximizes the SINR:
    \begin{equation}
        \u_i=\nu_{\max}\left(\H_{ii}\v_i\v_i^H\H_{ii}^H,\sum_{j\neq i}\H_{ij}\v_j\v_j^H\H_{ij}^H+\sigma^2{\bf I}\right),
    \end{equation}
    where $\nu_{\max}({\bf A},{\bf B})$ denotes the generalized eigenvector of the matrix pencil $({\bf A},{\bf B})$ with maximum generalized eigenvalue, and ${\bf I}$ is an identity matrix with the appropriate dimensions.
    \item Keeping the decoders fixed and changing the roles of transmitters and receivers, the precoders are obtained as those maximizing the SINR of the reversed communication, i.e.,
    \begin{equation}\label{eq:vO}
        \v_i=\nu_{\max}\left(\H_{ii}^H\u_i\u_i^H\H_{ii},\sum_{j\neq i}\H_{ji}^H\u_j\u_j^H\H_{ji}+\sigma^2{\bf I}\right).
    \end{equation}
    \item Steps 1 and 2 are repeated until convergence or until a prescribed number of iterations has been reached.
\end{enumerate}

For further details, we refer the reader to \cite{AltMinIA}.

\subsection{Interference Alignment with Pre-FFT Decoding}\label{subsec:IA_preFFT}
As mentioned in Section \ref{sec:synchronizationIssues}, the existence of symbol timing offsets between the desired and the interfering OFDM symbols impairs the synchronization procedure. Therefore, interference must be eliminated (or at least sufficiently reduced) in asynchronous scenarios before the synchronization step. To this end, pre-FFT IA decoders must be applied at the receiver side. Let us first consider a general time-domain spatial interference alignment approach in which both precoders and decoders are applied in time domain, ${\bf v}_j[n]\in\C^{2\times1},\ n = 0,\dots, L-1$, is the impulse response of the linear precoder with length $L$ for the transmitter $j$, and ${\bf u}_i[n]\in\C^{2\times1},\ n = 0,\dots, L-1$, is the impulse response of the pre-FFT linear decoder for receiver $i$, also with length $L$. The output signal at receiver $i$, $z_i$, is given by
\begin{align}
    z_i[n]=&\underbrace{{\bf u}_i^H[-n] \ast {\bf H}_{ii}[n] \ast {\bf v}_i[n] \ast x_i[n-\mu_{ii}]}_\text{desired link}+\notag \\
    &\underbrace{\sum_{j\neq i}{\bf u}_i^H[-n] \ast {\bf H}_{ij}[n] \ast {\bf v}_j[n] \ast x_j[n-\mu_{ij}]}_\text{multiuser interference}+\notag \\
    &\underbrace{{\bf u}_i^H[-n] \ast {\bf n}_i[n]}_{\text{noise}},
\label{Eq:received_signal_broadband}
\end{align}
where $n$ is the discrete-time sample index, $x_j[n]$ is the discrete-time OFDM signal transmitted by user $j$, ${\bf H}_{ij}[n]$ is the matrix impulse response of the frequency-selective MIMO channel between transmitter $j$ and receiver $i$, $\mu_{ij}$ denotes de delay between transmitter $j$ and receiver $i$, and $\ast$ denotes convolution. The received signal at user $i$ is also affected by an additive, spatially and temporally-white Gaussian noise ${\bf n}_i[n]\sim\mathcal{N}({\bf 0},\sigma^2{\bf I})$. Notice that we are now considering an asynchronous wireless system and, for this reason, a delay $\mu_{ij}$ is explicitly introduced in the signal model given by Eq.~\eqref{Eq:received_signal_broadband}.

As we already showed in \cite{ISWCS2012}, the interference leakage when precoding and decoding are both applied in the time domain is given by the sum of the energies of the equivalent interference channels, ${\bf u}_i^H[-n] \ast {\bf H}_{ij}[n] \ast {\bf v}_j[n]$ with $i\neq j$. In other words, the interference leakage is independent of the specific delays between users, $\mu_{ij}$, and hence this approach can work properly in the presence of symbol timing offsets. Note that for the interference to be mitigated before time synchronization, only time-domain decoders are strictly necessary, while precoders could be applied either in the time or in the frequency domain. Clearly, by precoding in the frequency domain, the interference leakage will depend on the delays between transmitters and receivers, and hence there will be some residual interference when the interfering symbols are not aligned in time with the receiver window. Nevertheless, this simple scheme makes time synchronization possible in the presence of asynchronous interferences and allows us to assess the performance degradation of time-domain decoding with respect to its frequency-domain counterpart.

Therefore, and for simplicity, we will consider that precoders operate in the frequency domain whereas the decoders are applied in the time domain (pre-FFT); and we propose in the ensuing lines a simple method to compute the pre-FFT decoders that mitigate the interference before time synchronization. Obviously, a pure time-domain approach with a specific design of the time-domain precoders and decoders, such as those proposed in~\cite{ISWCS2012,ICASSP2012}, would outperform the adopted approach but at the cost of an increased computation time for calculating the set of interference alignment precoders and decoders, thus impacting the feedback time. In any case, the design and evaluation of such approaches is beyond the scope of this paper.

A schematic of the pre-FFT IA decoding scheme is shown in Fig.~\ref{fig:SchematicPostPreFFT}~(bottom). Assuming again perfect CSI knowledge, we propose the following method for computing the pre-FFT IA decoders:
\begin{itemize}
    \item First, the IA precoders and decoders are computed on a per-subcarrier basis applying the closed-form solution described in Section~\ref{sec:ClosedFormIA}.
    \item Next, a $N_{\text{FFT}}$-point IFFT is applied to the set of post-FFT decoders in order to obtain their impulse response.
    \item Finally, the pre-FFT filters are truncated to a given length, $L$, so as to reduce the ISI and ICI.
\end{itemize}
Note that the shorter the impulse response of the equivalent channel ---consisting of the actual wireless channel convolved with the pre-FFT filters,--- the lower the ISI/ICI but the higher the residual Multi-User Interference (MUI), and vice versa. Thus, pre-FFT filtering involves a trade-off between both sources of interference \cite{ISWCS2012}.

\begin{figure}[t!]
\centering
\includegraphics[width=\myfigurewidth]{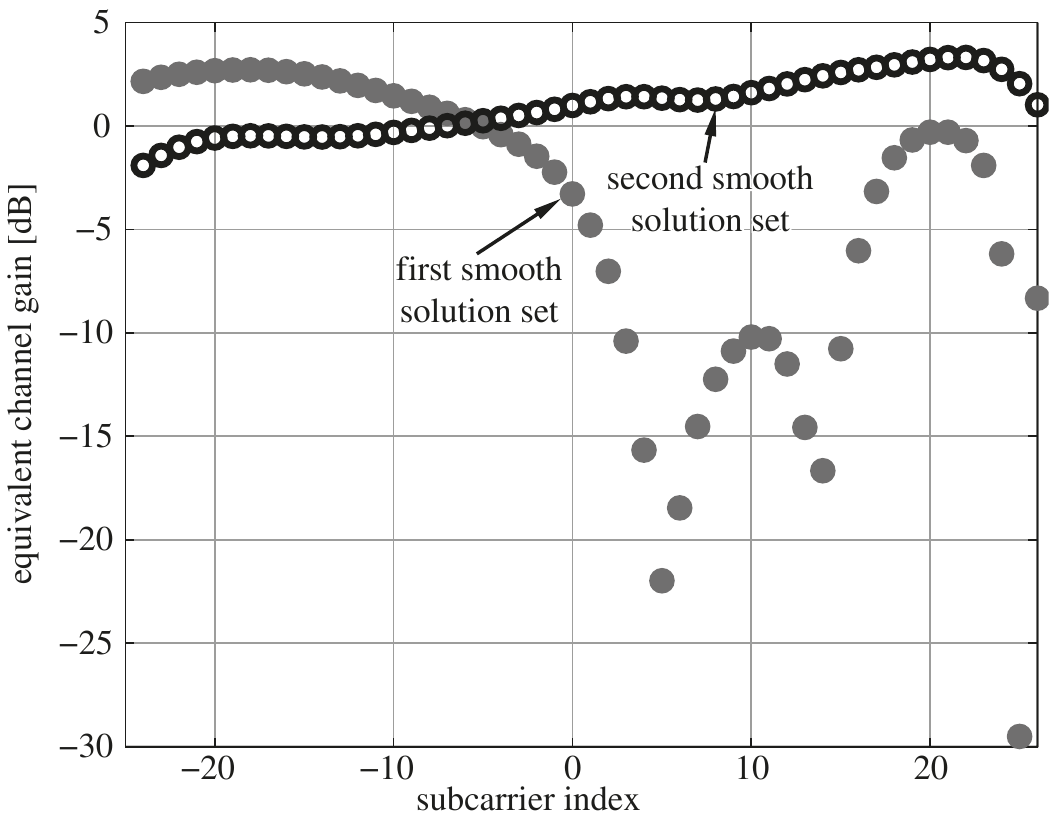}
\caption{{Example of the set of equivalent channels obtained with each solution in the frequency domain from simulated wireless channels.}
Notice that there are only two smooth sets out of $2^{N_{\text{FFT}}}$.}
\label{fig:IA_solutions}
\end{figure}

It is important to notice that the OFDM $(2\times2,1)^3$ interference channel is being interpreted as $N_{\text{FFT}}$ parallel single-carrier $(2\times2,1)^3$ interference channels and that there exist two IA solutions per subcarrier. Thus, there is a total of $2^{N_{\text{FFT}}}$ solutions in the OFDM case,\footnote{Notice that, in practice, the number of solutions will be noticeably lower due to the null subcarriers.} as each of the $N_{\text{FFT}}$ subcarriers can use a different solution without altering the alignment conditions. However, as we are interested in pre-FFT filters with an impulse response as short as possible (and, consequently, pre-FFT filters with a frequency response as smooth as possible in the frequency domain) it is important to select those solutions that provide the smoothest frequency response for the equivalent channel. Simulations have shown that there are only two sets of smooth solutions out of $2^{N_{\text{FFT}}}$. Figure~\ref{fig:IA_solutions} plots the magnitude of the frequency responses of one of the SISO equivalent channels obtained after calculating the IA precoders and decoders using one of these sets (empty circles) and the other one (full circles), respectively. Note that any other combination of these two sets leads to more abrupt changes in the frequency response.

\begin{figure}[t!]
\centering
\includegraphics[width=\myfigurewidth]{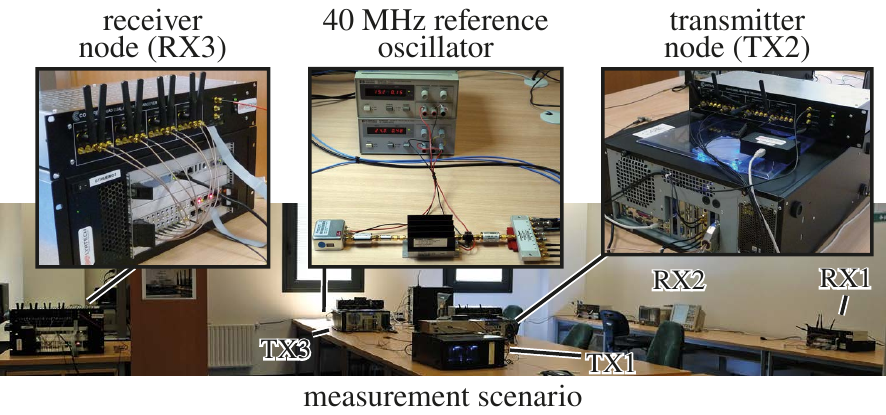}
\caption{{Picture of the measurement scenario.}
Detail (from left-hand to right-hand side) of a receiver node, the common frequency reference, and a transmitter node.}
\label{fig:testbed}
\end{figure}

\section{Multiuser MIMO Testbed}\label{sec:testbed}

This section describes the MIMO wireless network testbed that has been used to assess, in a realistic scenario, the IA techniques presented in the previous section. Both transmit and receive testbed nodes (see Fig.~\ref{fig:testbed}) have a Quad Dual-Band front-end from Lyrtech, Inc.~\cite{Lyrtech}. This RF front-end can use up to eight antennas that are connected to four direct-conversion transceivers by means of an antenna switch. Each transceiver is based on Maxim~\cite{Maxim} MAX2829 chip, which supports both up and down conversion operations from either 2.4 to 2.5\,GHz or 4.9 to 5.875\,GHz. The front-end also incorporates a programmable variable attenuator to control the transmit power value. The attenuation ranges from 0 to 31\,dB in 1\,dB steps, while the maximum transmit power declared by Lyrtech is 25\,dBm per transceiver.

The baseband hardware is also from Lyrtech. More specifically, each node is equipped with a VHS-DAC module and a VHS-ADC module, respectively, containing eight DACs and eight ADCs. Each pair of DAC/ADC is connected to a single transceiver of the RF front-end and the signals are passed in I/Q format.

Both transmit and receive nodes employ buffers that are dedicated to store the signals to be sent to the DACs as well as the signals acquired by the ADCs. The utilization of such buffers allows for the transmission and acquisition of signals in real-time, while the signal generation and processing is carried out off-line. Both baseband hardware and RF front-ends of the transmit nodes are synchronized in time and in frequency by means of two mechanisms:
\begin{itemize}
\item Transmit nodes implement a hardware trigger attached to the real-time buffers, and to the DACs and ADCs. When one of the nodes fires the trigger (usually the node corresponding to user $1$) for all buffers, DACs and ADCs receive the signal and start transmission and acquisition simultaneously (the timing between nodes is precise up to 2 samples, hence resulting in an error upper bound of $\pm$ 50\,ns).

\item The sampling frequency of DACs and ADCs is set to 40\,MHz, while the RF front-ends support a reference frequency of 40\,MHz. In order to synchronize all nodes in frequency, the same common external 40\,MHz reference oscillator is distributed to the DACs, ADCs and RF front-ends of all nodes, hence guaranteeing high-quality frequency synchronization.
\end{itemize}

The core component of each node is a host PC which allocates, configures and controls the baseband hardware and the RF front-end. Furthermore, the host PC provides remote control functionalities that allow the node to be externally controlled through socket connections. This flexible design has been found very useful because each node can be transparently controlled. Also, it allows a so-called control PC with standard TCP/IP connections to use Matlab to interact with the whole testbed, which considerably enhances the development of multiuser experiments. Moreover, this control PC acts as a feedback channel to share CSI among nodes and carries out all signal processing operations. The web page of the COMONSENS project \cite{COMONSENSweb} contains detailed information about the technical features of the testbed.

\begin{figure}[t!]
\centering
\includegraphics[width=\myfigurewidth]{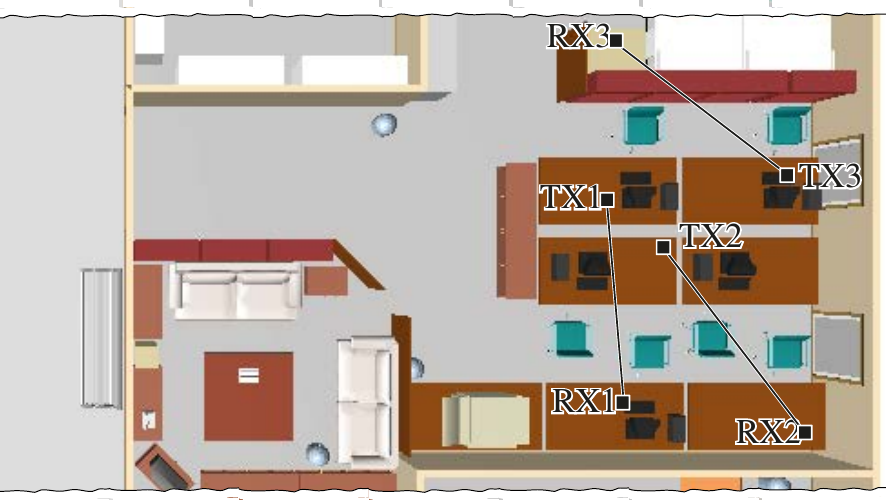}
\caption{{Plan of the measured scenario.}
The distances between users 1, 2 and 3 are approximately 3, 3.4, and 3.2 meters, respectively. Notice that the line-of-sight of the link between TX3 and RX3 is blocked by shelves.}
\label{fig:scenario}
\end{figure}

\section{Measurement Setup}\label{sec:setup}
\label{sec:measurement_setup}
Figure \ref{fig:scenario} shows the measurement scenario set up at the University of Cantabria to recreate a typical 3-user indoor interference channel. The access to the room was carefully controlled during the measurements to ensure that there were no moving objects in the surroundings. Additionally, we also checked that no other wireless system was operating in the 5\,GHz frequency band. All nodes were equipped with monopole antennas at both transmitter and receiver sides \cite{MobileMark,Vert2450}, while the antenna spacing was set to approximately seven centimeters (forced by the separation of the antenna ports at the RF front-end).

\begin{figure}[!t]
\centering
\includegraphics[width=\myfigurewidth]{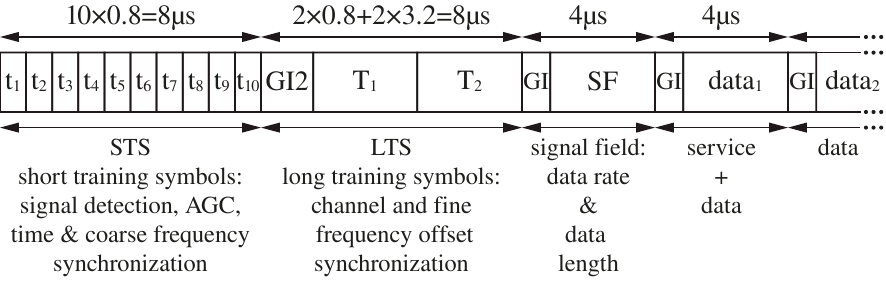}
\caption{{IEEE 802.11a physical-layer frame.}
}\label{fig:StandardFrame}
\end{figure}

\begin{figure}[!t]
\centering
\includegraphics[width=\myfigurewidth]{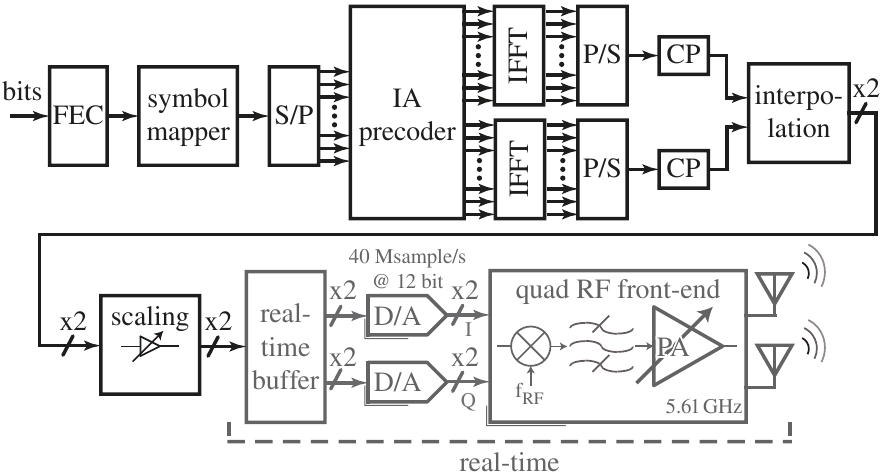}
\caption{{Block diagram of hardware and software elements at each transmit node.}
IA precoding is performed before the IFFT.}
\label{fig:TXTestbed}
\end{figure}

We followed the IEEE 802.11a WLAN standard with the conventional frame structure and synchronization headers. Figure \ref{fig:StandardFrame} shows the 802.11a physical-layer frame. The header comprises the Short Training Sequence (STS), the Long Training Sequence (LTS), and the Signal Field (SF). The STS is used for frame detection, while the LTS is utilized for frequency offset correction and channel equalization. The SF contains information about the data rate and the frame length. During our experiments, the frame length remained constant and there was no rate adaptation. Therefore, we made no use of the information conveyed by the SF.

Each OFDM symbol contains 48 data symbols and 4 pilots, which were OFDM-modulated using a 64-point IFFT. The CP length is 16 samples (800\,ns).

\begin{table*}[t]
\begin{center}
\begin{tabular}{cccccc}
\toprule
\multicolumn{4}{c}{TRAINING FRAMES ($\mathring{z}_{s,n}=\mathring{z}_{s}\;\forall n$)} & DATA FRAMES \\
\cmidrule(r){1-4} \cmidrule{5-5}
Channel & Signal power & Residual noise power & Residual noise variance & EVM \\
\cmidrule(rl){1-1}\cmidrule(rl){2-2}\cmidrule(rl){3-3}\cmidrule(rl){4-4} \cmidrule(rl){5-5}
$\displaystyle h_s=\frac{\sum_nz_{s,n}}{M\mathring{z}_s}$ &
$\displaystyle  S_s=\left|\frac{\sum_nz_{s,n}}{M}\right|^2$ &
$\displaystyle N_s=\frac{\sum_n\left|z_{s,n}-h_s\mathring{z}_s\right|^2}{M}$ &
$\displaystyle \sigma^2_s=\frac{N_s}{\left|h_s\right|^2}$ &
$\displaystyle \text{EVM}_s=\frac{\sum_n\left|\bar{z}_{s,n}-\mathring{z}_{s,n}\right|^2}{\sum_n\left|\mathring{z}_{s,n}\right|^2}$ \\
\midrule
\multicolumn{5}{c}{$\begin{array}{ccc}
z_{s,n}\text{: received symbol}&  \bar{z}_{s,n}\text{: equalized received symbol} & \mathring{z}_{s,n} \text{: transmitted symbol}
\end{array}$}\\
\multicolumn{5}{c}{$\begin{array}{cc}
s\text{: subcarrier index} & n\text{: OFDM data symbol index (from 1 to N)}\\
M\text{: number of OFDM training symbols} & N\text{: number of OFDM data symbols}
\end{array}$}\\
\bottomrule
\end{tabular}
\caption{Formulas applied to training or data frames for obtaining the different parameters.}
\label{tab:formulas}
\end{center}
\end{table*}

Figure \ref{fig:TXTestbed} shows the general block diagram of a transmitter chain which differs from that of a conventional 802.11a in the IA precoding block and in the utilization of two transmit antennas. Both software and hardware elements perform the following steps:
\begin{itemize}
	\item The source bits are encoded (convolutional code, scrambling and interleaving) and mapped to a BPSK, QPSK, 16-QAM, or 64-QAM constellation (see ``FEC'' and ``symbol mapper'' blocks in Fig.~\ref{fig:TXTestbed}) depending on the transmission rate according to the 802.11a standard (cf.~\cite{80211a}). The data frame length is set to a reasonable length for a WLAN frame (1250 bytes), which depending on the transmission rate translates into a different number (denoted by $N$ in Table~\ref{tab:formulas}) of OFDM data symbols.
	\item An IA precoder is applied to each subcarrier and two OFDM symbols (one for each antenna) are generated.
	\item At each antenna, the OFDM-sampled symbols are encapsulated into 802.11a standard-compliant frames and afterwards they are upsampled by a factor of two.
	\item The resulting signals are scaled so as to have a mean transmit power of 5\,dBm per antenna, quantized according to the 12-bit resolution of the DACs, and finally stored in the real-time buffers available at the transmit nodes of the testbed.
    \item At this point, the transmitters are ready to receive the trigger signal and start transmitting simultaneously. Once the transmit nodes are triggered, the buffers containing the OFDM signal are read by the corresponding DACs at a rate of 40\,Msample/s. Next, the analog signals are sent to the RF front-end in order to be transmitted at the center frequency of 5610\,MHz. Simultaneously, the three receivers start to acquire the transmitted frames.
\end{itemize}

\begin{figure}[!t]
\centering
\includegraphics[width=\myfigurewidth]{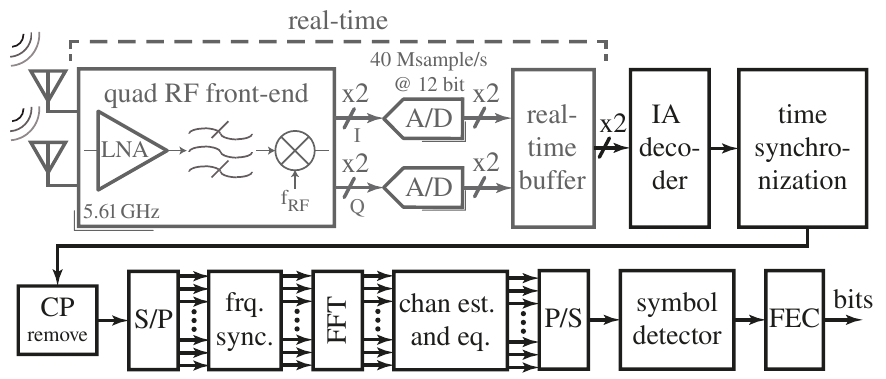}
\caption{{Block diagram of hardware and software elements at each receive node.}
IA decoding is performed before the FFT.}
\label{fig:RXTestbed}
\end{figure}

Figure~\ref{fig:RXTestbed} shows the hardware and software elements corresponding to a receiver implementing pre-FFT decoding. Notice the position of the IA decoder block, which is applied in time domain before synchronization takes place, as explained in Section~\ref{subsec:IA_preFFT}. The pre-FFT IA decoder generates one stream of data that is subsequently processed following a typical 802.11a receiver structure. The block diagram corresponding to post-FFT decoding is analogous to the transmitter shown in Fig.~\ref{fig:TXTestbed} and does not require an additional description.

The trigger signal is received by both transmitters and receivers simultaneously. When triggered, the receive nodes carry out the following operations:
\begin{itemize}
	\item The RF front-end down-converts the signals received by the selected antennas to the baseband, generating the corresponding I and Q analog signals.
	\item All I and Q signals (four in total) are then digitized by the ADCs at a sample rate of 40\,Msample/s and they are stored in the real-time buffers.
	\item {The I and Q signals are decimated by a factor of two.}
	\item The signals are properly scaled according to the 12 bits ADC resolution. Notice that this factor is constant during the course of the whole measurement, thus not affecting the properties of the wireless channel.
	\item (Only for pre-FFT decoding) The acquired signals are processed by the pre-FFT IA decoder which generates a single data stream to be processed by a standard 802.11a receiver.
	\item Frame detection and time synchronization takes place.
	\item The frame is properly disassembled and the OFDM symbols are parallelized and synchronized in frequency.
    \item The 64-point FFT is applied. Only once for pre-FFT decoded frames and twice for post-FFT decoded frames. Note that if the IA decoder is applied at the frequency domain (post-FFT), then the signals coming from the two receive antennas are processed separately, including the FFT operation, up to the point in which the IA decoding is applied.
    \item (Only for post-FFT decoding) Frequency-domain symbols are processed by the post-FFT IA decoder which generates a single data stream for the next processing blocks.
    \item The next step is least squares channel estimation and zero-forcing equalization.
	\item Finally, a symbol-by-symbol hard decision decoding is performed followed by a channel decoder which outputs the estimated transmitted bits.
\end{itemize}

\section{Measurement Methodology}\label{sec:methodology}

Success in the experimental evaluation of wireless communication systems relies mainly on the utilized measurement methodology, which depends on the scenario and the methods to be assessed. Given the complexity of the setup (see Fig.~\ref{fig:IAScheme}) the correct design of the measurement methodology is even more critical. In order to perform a fair comparison, it is necessary that the measurement methodology supports the assessment of several figures of merit, with and without interference, while guaranteeing that in both cases the signals experience the same channel realization. The methodology should also allow us to measure the amount of interference created by each user as well as the interference leakage.

The proposed measurement methodology consists of two stages that require two different OTA signal transmissions for the assessment of a single frame per user. The first one is termed training stage because its objective is to obtain an estimate of the nine $2\times2$ MIMO channels of the 3-user interference channel. Once all channel estimates are available, the precoders and decoders of the different adopted schemes are computed and the second stage takes place. Aligned signals as well as signals from other schemes are sent ---in a single transmission cycle--- during this second stage in order to evaluate the performance of the IA approach and to compare such performance to that exhibited by other alternative approaches, all of them experiencing the same channel realization (notice that the wireless channel can be estimated free of interference and in an independent way for each transmission scheme in order to verify that all schemes experienced the same channel realization).

\begin{figure}[!t]
\centering
\includegraphics[width=\myfigurewidth]{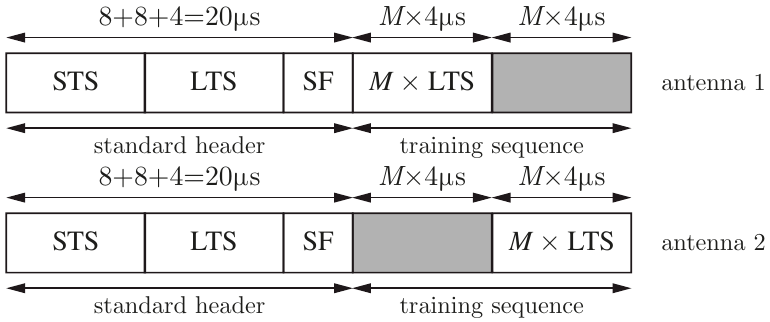}
\caption{{Training frame for MIMO channel estimation.}
For each transmit antenna, we use a training sequence comprised of $M$ long-training OFDM symbols.}
\label{fig:TrainingFrame}
\end{figure}

To conduct the training stage, we have introduced the training frame shown in Fig.~\ref{fig:TrainingFrame}, whose format differs from that of the IEEE 802.11a standard. In the following, we detail the measurement procedures performed at each stage.

\begin{itemize}
    \item \textbf{Training stage}: all users sequentially send over each transmit antenna training frames comprised of $M$ long-training OFDM symbols in a TDMA fashion (only a single user is transmitting at a given time instant), while the three receivers are simultaneously acquiring. Once the training signals have been acquired, the nine pairwise MIMO channels are estimated and the precoders and decoders for each transmission scheme are computed.
    \item \textbf{Data transmission stage}: users transmit data frames comprised of $N$ OFDM symbols according to different transmission schemes. Signals corresponding to the following schemes are sent one after each other (without delays between them):
        \begin{enumerate}
            \item \textbf{IA transmission}: all users transmit simultaneously, hence creating a 3-user interference channel. The IA precoders are applied at the transmitter in the frequency domain right before the FFT.
            \item \textbf{Perfect IA transmission}: each user applies the same set of precoders as in the previous scheme, next the resulting signals are transmitted in a sequential fashion, i.e., from only one user at a time. This transmission scheme enables us to measure the residual interference level created by each transmitter at each receiver. In other words, we are able to evaluate the impact of the residual interference by comparing the actual performance during the IA stage with that in the absence of interference.
            \item \textbf{MaxSINR transmission}: all users transmit simultaneously, creating again a 3-user interference channel. The IA precoders and decoders are computed with the MaxSINR algorithm, as explained in Section~\ref{sec:MaxSINR}. The noise variance has been obtained according to Table~\ref{tab:formulas}, and the algorithm has proceeded until convergence with a random initial point. While IA focuses exclusively on canceling the interference without paying attention at the quality of the resulting equivalent channel, MaxSINR trades interference mitigation and desired signal enhancement, which may provide a performance improvement if the SNR is not sufficiently high for IA to be optimum.
            \item \textbf{DET-TDMA transmission}: users transmit sequentially through the principal eigenvectors of the channel. This scheme is sometimes denoted as Dominant Eigenmode Transmission (DET)~\cite{Andersen} and provides the best equivalent channel response. Therefore, it allows us to evaluate the degradation of the desired links when all available antennas are entirely employed for interference mitigation.
            \item \textbf{SISO-TDMA transmission}: users transmit sequentially using a single antenna for transmission and reception, hence creating a standard-compliant 802.11a link. In the experiments, each transmitter uses the first antenna while both antennas are sequentially used for reception. This strategy provides data transmitted over two different SISO channel realizations and more accurate results after averaging.
        \end{enumerate}
\end{itemize}

For each channel realization, the foregoing procedure is repeated for all individual data rates specified by the IEEE 802.11a standard. Therefore, a training stage followed by a data transmission stage is conducted for each data rate. Notice that the Medium Access Control (MAC) layer in the IEEE 802.11a standard adapts the data rate according to the quality of the received signal. In our experiments, however, we fix the rate regardless of the reception quality.

\begin{figure}[!t]
\centering
\includegraphics[width=\myfigurewidth]{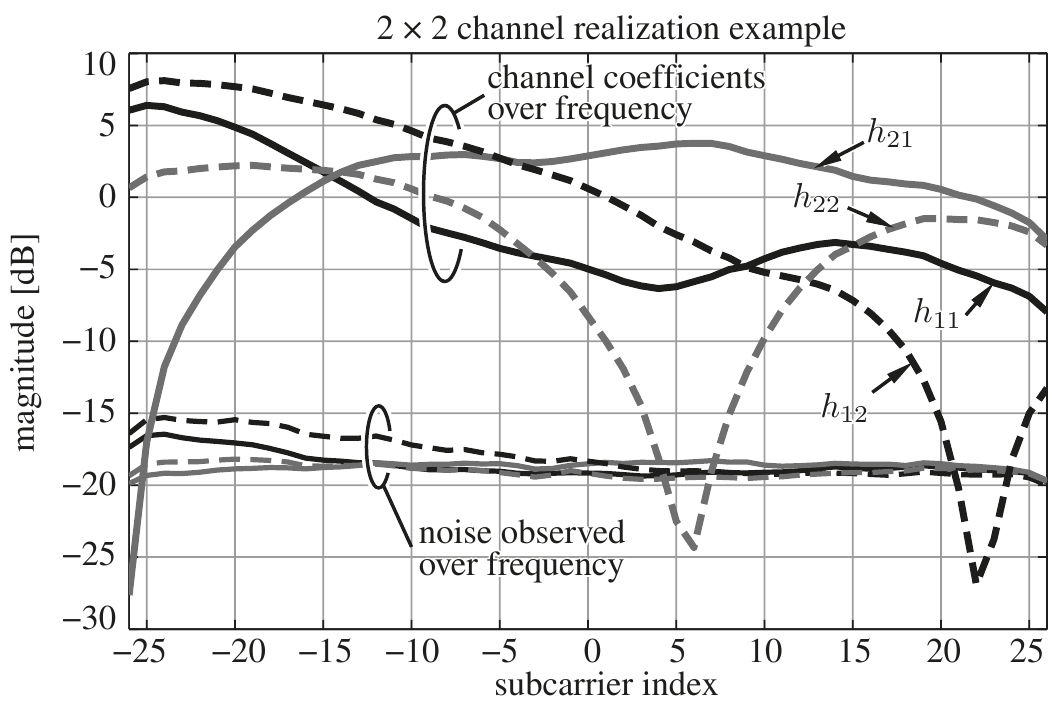}
\caption{{Example of one of the $2\times2$ MIMO channels obtained in the measurement scenario.}
The four noise estimates that were obtained using the four SISO links are also depicted in the figure. Notice that the noise power is not flat over frequency and varies according to the amplitude of the corresponding channel coefficient.}
\label{fig:ChannelExample}
\end{figure}

\section{Results}\label{sec:results}

\subsection{Characterization of the Channel Realizations}
\label{sec:channel_realizations}
In order to ensure statistically representative results, we conducted a sufficiently large number of executions of the aforementioned procedure over different wireless channels. In particular, binary switches allowed us to choose four different two-antenna sets at each node which makes a total of 4096 different channel realizations (all channels are available for download in the web page of the COMONSENS project \cite{COMONSENSweb}). Note that the corresponding interference alignment solution is completely different as long as a single channel coefficient of the three channel matrices changes. We recall that the position of the nodes nor the transmission frequency were changed.

First of all, we characterized the quality of the channels in our setup. In Fig.~\ref{fig:ChannelExample} we provide an example of the frequency response magnitude (normalized by the average of the channel amplitudes) for one of the measured $2\times2$ MIMO channels. Figure~\ref{fig:ChannelExample} also plots the estimated noise power at each receive antenna obtained as indicated in Table~\ref{tab:formulas}. Notice that we can obtain four noise variance estimates, one for each transmit-receive antenna pair. It can be observed that the noise level is not flat over frequency and follows the quality of the corresponding channel coefficient, i.e., it is proportional to the channel gain. This behavior is explained by signal distortion at the transmitter, also referred to as transmitter noise \cite{TXnoiseEURASIP2012}. Hereinafter, we will refer to this estimated noise as residual noise according to the way it is calculated.

\begin{figure}[!t]
\centering
\includegraphics[width=\myfigurewidth]{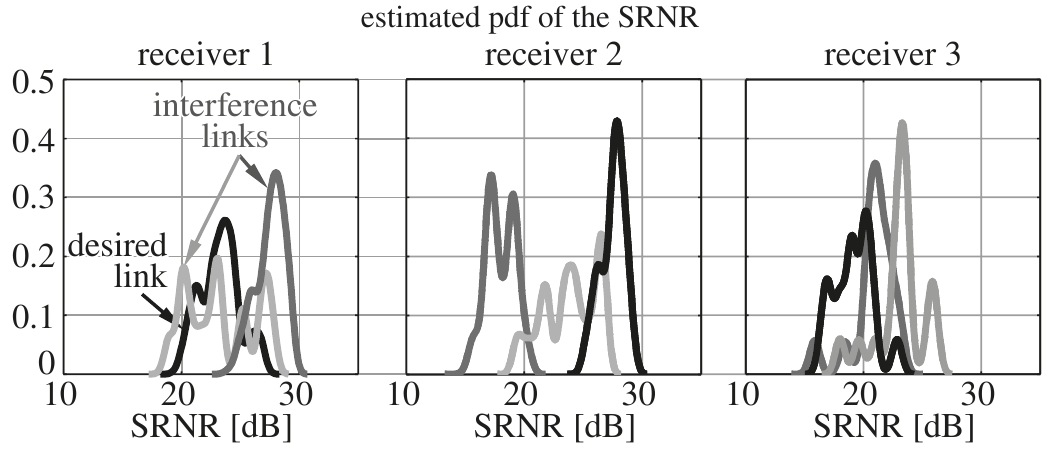}
\caption{{Estimated PDF of the SRNR of the desired and interfering links at each receiver.} The SRNR is estimated during the training stage according to the formulas given in Table~\ref{tab:formulas}. More specifically, the SRNR is computed for each subcarrier index and averaged over all OFDM symbols in a given frame. Finally, the PDF is estimated from the obtained set of SRNR values.}
\label{fig:fdp_SNR}
\end{figure}

Then, we define the Signal-to-Residual Noise Ratio (SRNR) as the ratio between the estimated signal power and the residual noise power. Notice that the SRNR serves as a pessimistic proxy for the SNR as it accounts for the combined effect of the thermal noise at the receiver and the signal distortion at the transmitter. Figure~\ref{fig:fdp_SNR} shows the estimated Probability Density Function (PDF) of the SRNR at each receiver for both the desired and the interfering links. The SRNR for each subcarrier has been obtained with the expressions indicated in Table~\ref{tab:formulas}. As shown in the figure, the SRNRs range from approximately 15 to 30\,dB, with significant differences among receivers: interference is slightly stronger than signal at receivers 1 and 3, whereas receiver 2 experiences higher signal strength.

These measurements demonstrate the suitability of the scenario for the evaluation of IA techniques: first, all desired and interfering signals are of comparable strength and, second, SRNRs are relatively high.

\subsection{Comparison of Pre-FFT and Post-FFT IA Decoding}
\label{sec:comparison_preFFT_postFFT}

\begin{figure}[!t]
\centering
\includegraphics[width=\myfigurewidth]{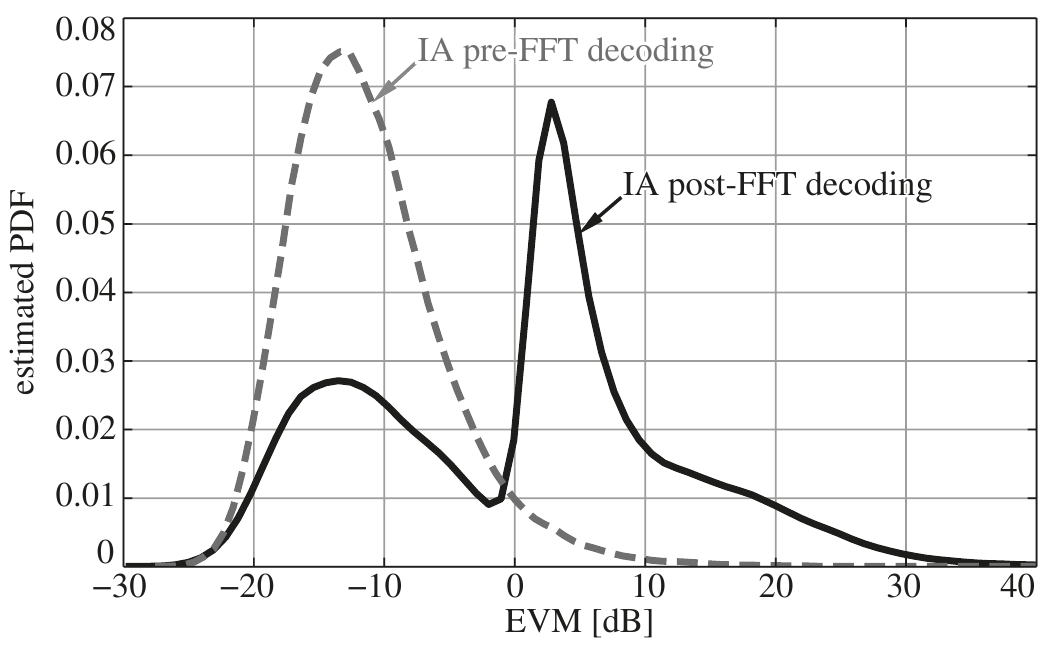}
  \caption{{Estimated PDF of the received constellation EVM for both pre- and post-FFT decoding under asynchronous transmissions {($M=30$, $L=64$, 16-QAM)}.}
      The asynchronous transmissions have been emulated using the received signals during the Perfect IA transmission stage.}
      \label{fig:AsynchrEVM}
\end{figure}

\begin{figure}[!t]
\centering
\includegraphics[width=\myfigurewidth]{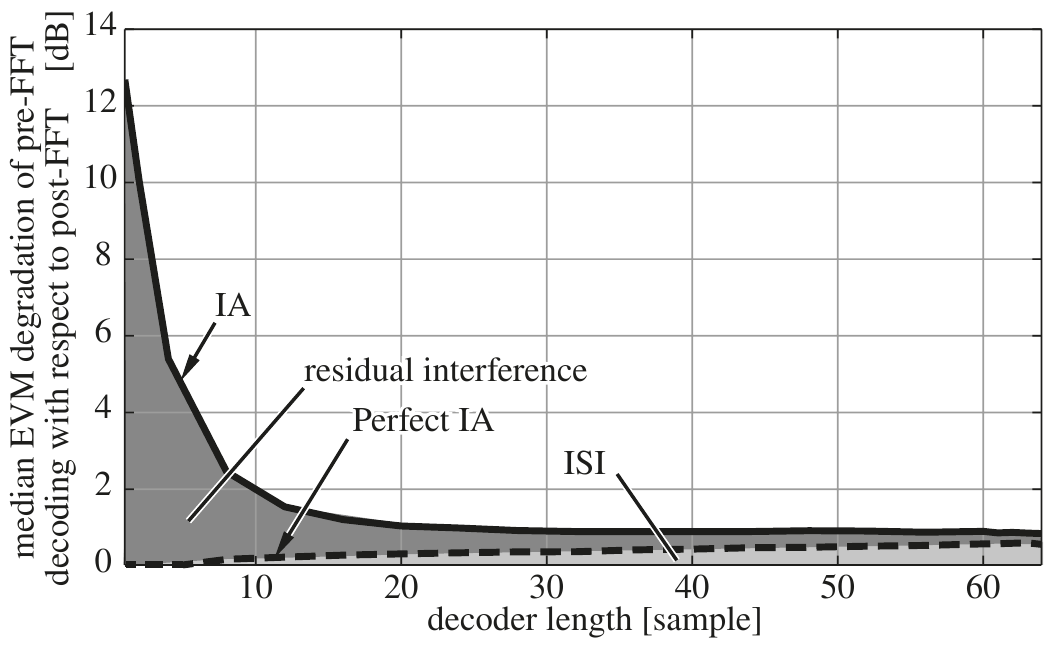}
\caption{{EVM degradation of time-domain (pre-FFT) decoded OTA transmissions with respect to the frequency-domain (post-FFT) counterpart {($M=30$, 16-QAM)}.}
The EVM degradation is plotted for both Perfect IA transmission (absence of interference) and IA transmission (presence of interference). Notice that a 0\,dB degradation means that both pre-FFT and post-FFT decoding methods perform the same in terms of EVM.}
\label{fig:EVMdegradation_vs_decoderLength}
\end{figure}

In this section we experimentally evaluate the performance of the pre-FFT (time-domain) IA decoding scheme proposed in Section~\ref{subsec:IA_preFFT} in comparison to post-FFT (frequency-domain) decoding while at the transmitter IA precoding is applied in the frequency domain and the three transmitters are synchronized among them. All OTA transmissions are carried out at 24\,Mbit/s (16-QAM) and the EVM of the received signal constellation (calculated as in Table~\ref{tab:formulas}) is used as the performance metric.

We start assessing the performance of both decoding schemes in an asynchronous scenario where users transmit at arbitrary time instants. Notice that, although all measurements have been carried out under synchronized transmissions, we can emulate an asynchronous scenario by using the received signals during the Perfect IA transmission stage. Indeed, in the Perfect IA transmission stage, each receiver node acquires an interference-free version of the signal transmitted by each user. Before processing these frames, an arbitrary delay can be included in each of the received signals, which are then summed up to yield a received signal comprised of the three time-misaligned frames. Note that the resulting SRNR will be approximately 4.7\,dB lower, since the noise is also added up in this process. Nevertheless, the relative performance of pre-FFT and post-FFT IA decoding will not be noticeably affected by such SRNR reduction. We plot in Fig.~\ref{fig:AsynchrEVM} the estimated PDF of the received constellation EVM for $M=30$ training symbols and a decoder length of $L=64$ samples. It can be observed that the post-FFT decoding scheme yields a bimodal PDF, which is the result of the aforementioned synchronization issues inherent to the post-FFT approach. When the time and frequency synchronization stages work properly, the resulting EVM is similar to that of the pre-FFT scheme. However, Fig.~\ref{fig:AsynchrEVM} shows that there is a large number of channel instances where synchronization fails, as revealed by the second mode of the estimated PDF. On the other hand, the pre-FFT IA decoding scheme avoids this situation, hence substantially outperforming its post-FFT counterpart in asynchronous scenarios.

\begin{figure}[!t]
\centering
\includegraphics[width=\myfigurewidth]{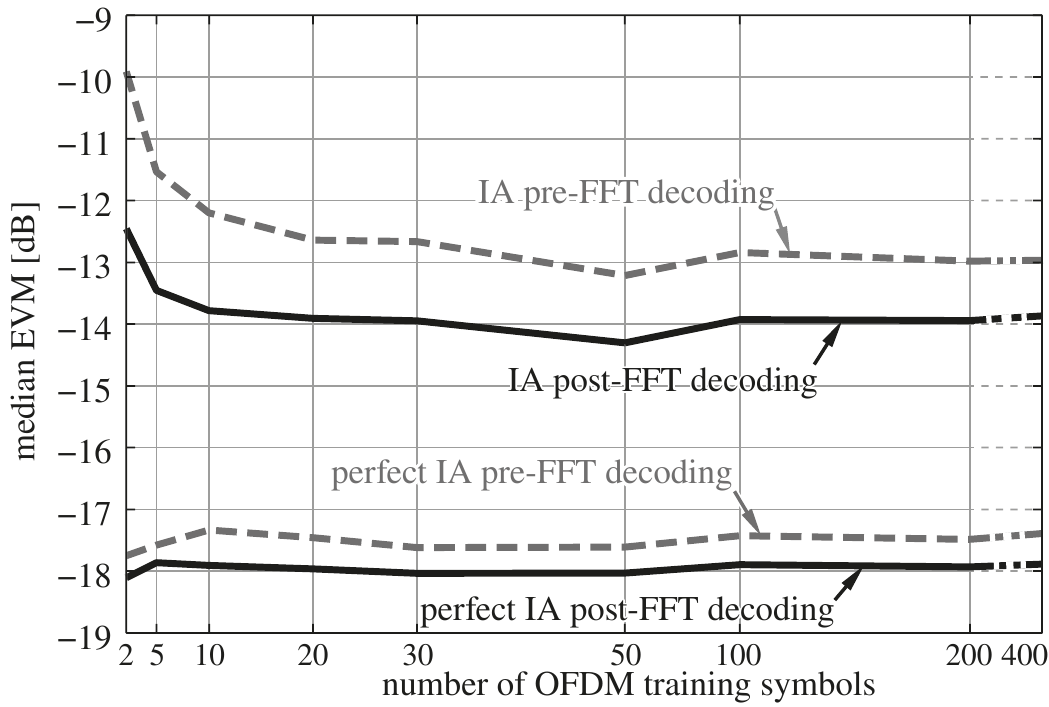}
\caption{{Evaluation of the median constellation EVM with the number of OFDM training symbols used in the training stage {($L=30$, 16-QAM)}.}
}\label{fig:EVM_vs_Npilots}
\end{figure}

To further evaluate the pre-FFT IA decoding scheme, we will focus on the synchronized setting in the remaining of this section. We first study the impact of the pre-FFT decoder length on the performance of IA which, as mentioned in Section~\ref{subsec:IA_preFFT}, involves a trade-off between ISI and residual MUI. To this end, we evaluate the EVM of the received signal constellation when MUI is suppressed with both post- and pre-FFT decoders. Training frames consist of $M=30$ training OFDM symbols per transmit antenna. Figure~\ref{fig:EVMdegradation_vs_decoderLength} shows the median EVM degradation of the pre-FFT technique for different decoder lengths, $L \in [1,64]$, with respect to the post-FFT decoder which obviously provides the best performance. In order to demonstrate the ISI \textit{versus} residual MUI trade-off, the comparison has been carried out for both IA and Perfect IA transmissions. For Perfect IA, the degradation is only due to ISI and, as expected, it increases with the decoder length. On the other hand, a shortened IA decoder cannot properly suppress the MUI leading to a high degradation of the constellation EVM. As the decoder length increases, however, the amount of MUI is greatly reduced whereas the degradation due to ISI grows at the rate seen in the Perfect IA curve. This analysis illustrates the existing ISI-MUI trade-off from which it turns out that a good choice for the decoder length would be 30 taps. This decoder length will be used in the remaining experiments since it provides slightly less than 1 dB of EVM degradation (whereof around 0.3 dB are due to ISI) with the advantage of a reduced receiver complexity and the possibility to perform frame synchronization in totally unsynchronized scenarios (as revealed in Fig.~\ref{fig:AsynchrEVM}).

\begin{figure}[!t]
\centering
\includegraphics[width=\myfigurewidth]{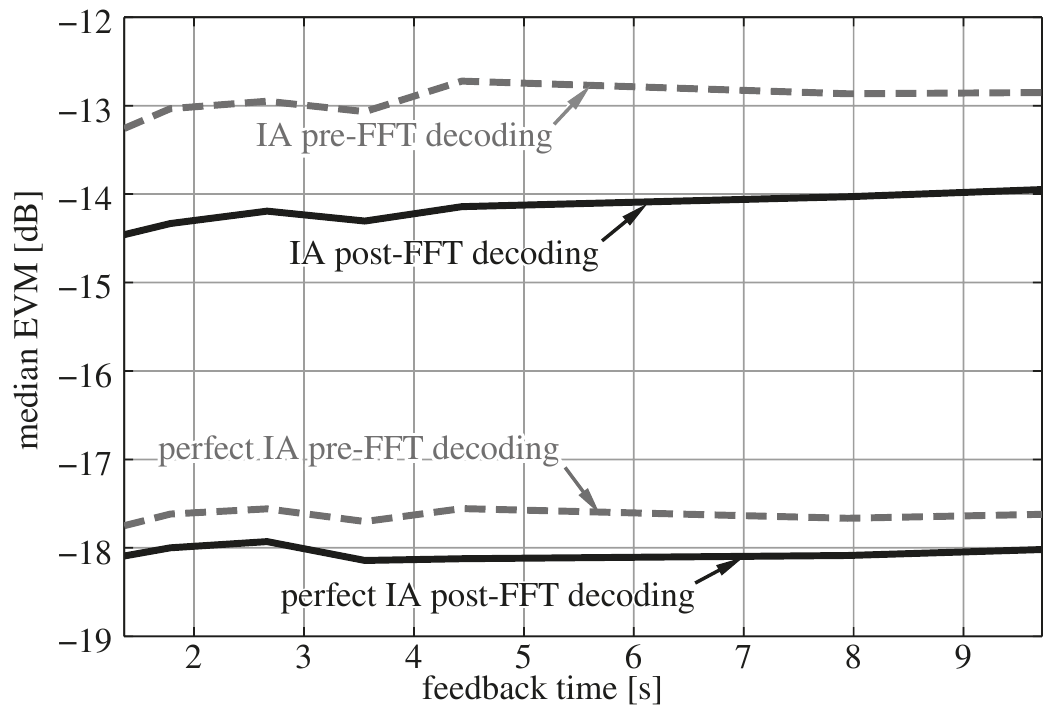}
\caption{{Evaluation of the median constellation EVM with the feedback time {($M=30$, $L=30$, 16-QAM)}.}
The feedback times are intentionally introduced and are ranging from 1 to 10 seconds, approximately.}
\label{fig:EVM_vs_feedbackTime}
\end{figure}

Secondly, we evaluate the effect that the quality of the CSI has on the performance of aligned transmissions considering our setup. In Fig.~\ref{fig:EVM_vs_Npilots} we show the evolution of the EVM for different number of OFDM training symbols. From the two upper curves (corresponding to IA transmissions) it can be observed that a small number of training symbols, below 20 or 30, does not provide an accurate CSI and leads to a significant degradation of the EVM due to interference. On the other hand, a number of training symbols above 30 does not improve the EVM anymore, which leads to a constant degradation between Perfect IA and IA of around 4 dB. The fact that the EVM does not improve when increasing the number of training symbols suggests that the performance of IA is not only limited by imperfect CSI but also by other spurious effects{, such as those derived from reusing the same training sequence and thus exciting the same nonlinearities each time.} Notice also that the gap between post-FFT and pre-FFT decoding is substantially higher for IA than for Perfect IA. This is due to the fact that, for the latter case, the degradation between both decoding schemes is caused by the additional ISI introduced by the pre-FFT decoding process (since the MUI has been avoided by sequential transmissions), whereas for the former case is due to ISI and residual MUI (see Fig.~\ref{fig:EVMdegradation_vs_decoderLength}). {Additionally, the impact of transmitter noise is also lower for Perfect IA since a single user is transmitting at a time instead of three simultaneously, as in the case of pre-FFT or post-FFT IA.}

One possible source of degradation could be the channel variations between the training stage and the data transmission stage, since the channel estimates used to compute the IA precoders and decoders are outdated by the time the aligned precoded transmission is actually performed. To evaluate this hypothesis, we conducted an additional experiment where a deliberate feedback time was introduced\footnote{Note that we do not intend to study the performance of IA with respect to the feedback time in general. We want to prove that our results are not affected by the feedback time required by our measurements (about a second).}. The results in Fig.~\ref{fig:EVM_vs_feedbackTime} show that increasing feedback time does not cause additional degradation of the received signal EVM, hence proving the channel remains static for at least 10 seconds. {Notice, however, that the performance could improve for feedback
times shorter than a second, which we cannot measure.} This is consistent with the special care taken to guarantee that our measurement scenario is completely static (see Section~\ref{sec:measurement_setup}). From the results shown in Figure~\ref{fig:EVM_vs_feedbackTime} we can ensure the validity of our measurement methodology regardless of the feedback time.

Once both the hypotheses of having inaccurate and outdated CSI estimates have been ruled out, there are still other reasonable effects which may jointly limit the performance of IA and may not completely disappear when using a large number of pilot symbols.

\begin{figure}[!t]
\centering
\includegraphics[width=\myfigurewidth]{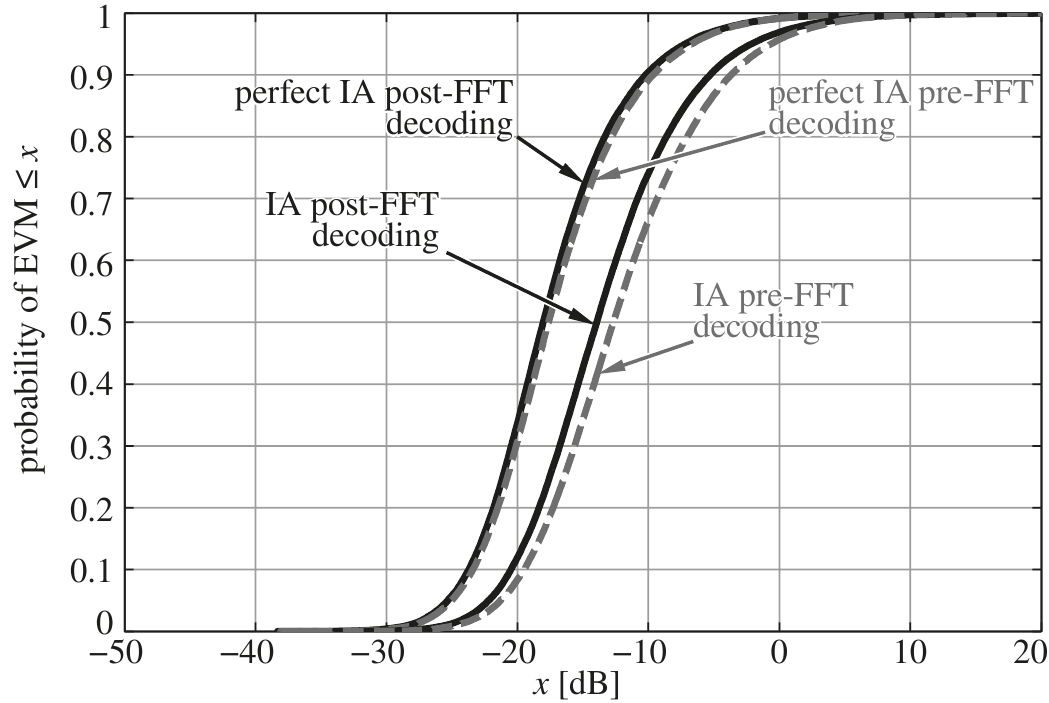}
\caption{{Comparison of the CDF of the received constellation EVM for both pre-FFT and post-FFT decoding {($M=30$, $L=30$, 16-QAM)}.} IA precoding is applied at the transmitter in the frequency domain on a per-subcarrier basis. $M=30$ training symbols and a decoder length of $L=30$ samples are used.}
\label{fig:cdf_EVM_preFFT_vs_postFFT}
\end{figure}

\begin{figure}[!t]
\centering
\includegraphics[width=\myfigurewidth]{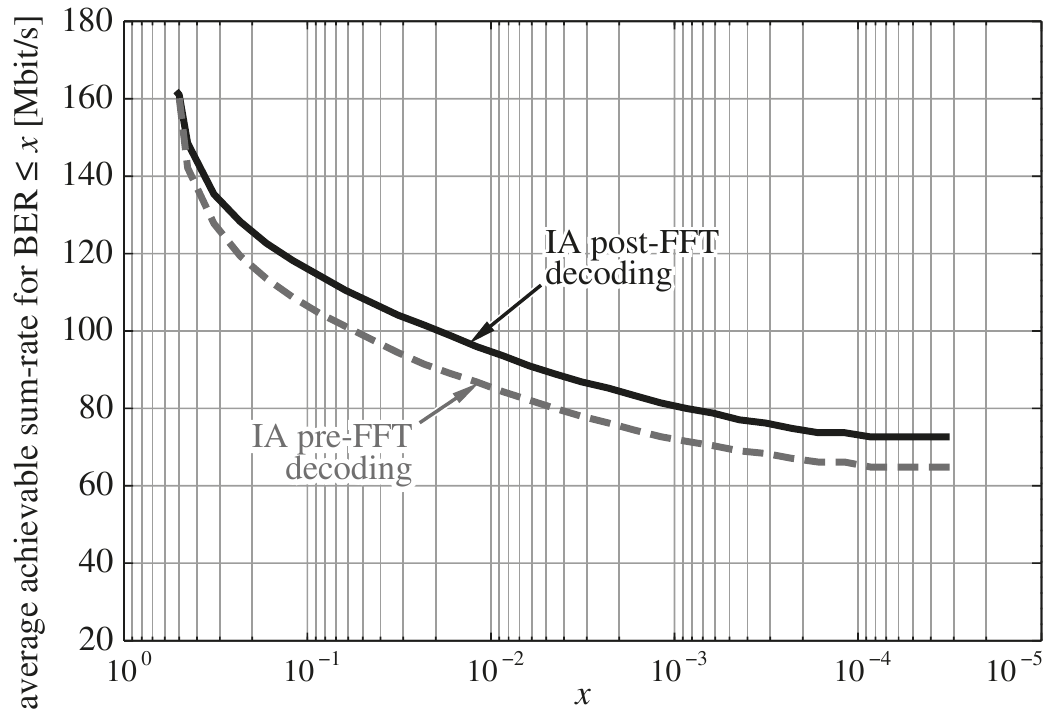}
\caption{{Average achievable sum-rate that guarantees a given BER for pre-FFT and post-FFT decoding methods {($M=30$, $L=30$)}.}
}\label{fig:BERvsSR_PreVsPostFFT}
\end{figure}

The impact of non-linearities in power amplifiers and RF oscillator phase noise on IA was empirically evaluated in \cite{Zetterberg2011}. When the signal distortion occurs at the transmitter it is known as transmitter noise and leads to spatially colored noise at the receiver. Transmitter noise, also referred to as dirty RF, is specially important when the transmitter and the receiver are close to each other since its effect is directly proportional to the channel power gain. Section~\ref{sec:channel_realizations} shows that transmitter noise is present in our measurement campaign. Its detrimental impact on the performance of MIMO systems is already well-known and has been empirically studied in \cite{TXnoiseEURASIP2012,suzuki2008,studer2010}. Some other effects, such as amplification gain drift (also known as transmitter droop) \cite{Walker2011} and packet-to-packet power variations, have not been studied yet in the context of IA. Notice that these effects may be specially pernicious for spatial-domain IA since they lead to power fluctuations at the transmitter over time which are different for every antenna and packet, and therefore cannot be fought with training.

A completely different explanation for the degradation stems from the fact of  applying precoding in the time domain at the transmitter side, i.e., on a per-subcarrier basis. As we pointed out in Section~\ref{sec:comparison_preFFT_postFFT}, the only way to suppress the interference independently of the delays between transmitters and receivers is to design both precoders and decoders to be applied in the time domain. In our experiments, as precoding is applied in the frequency domain at the transmitters, each receiver sees a residual interference that is proportional to the relative delay between the incoming desired and interfering signals. However, a careful experimental analysis of this issue and how largely it affects IA is still necessary and we leave it as a future work.

Finally, in view of the results in Figs.~\ref{fig:EVMdegradation_vs_decoderLength} and \ref{fig:EVM_vs_Npilots}, we have chosen the parameters which provide a nearly optimal performance with a reasonable complexity, that is, $M=30$ training symbols and a decoder length of $L=30$ samples. The Cumulative Distribution Function (CDF) of the received constellation EVM obtained with this parameter setup is shown in Fig.~\ref{fig:cdf_EVM_preFFT_vs_postFFT}. It is shown that the performance loss caused by moving from a frequency-domain to a time-domain decoder is always below 1\,dB for IA, and below 0.5\,dB for Perfect IA (while, in both cases, IA precoders are applied in the frequency domain at the transmitter). As a counterpart, time-domain IA decoding has the advantage that no inter-user time synchronization is required. Additionally, these differences are negligible compared to the roughly 4\,dB difference between Perfect IA and IA schemes shown in Fig.~\ref{fig:cdf_EVM_preFFT_vs_postFFT}.

Alternatively, we show BER results for both approaches in Fig.~\ref{fig:BERvsSR_PreVsPostFFT}. This figure represents the average achievable sum-rate that guarantees a BER equal to or lower than a given value. For each channel realization, the achievable sum-rate is obtained assuming an optimal MAC layer which selects for each user the maximum rate that satisfies the required BER. It is important to notice that results in Fig.~\ref{fig:BERvsSR_PreVsPostFFT} do not take additional overhead or higher-level issues into account and they only suggest how the optimum performance of such schemes would be. The difference of 1\,dB in terms of EVM between time-domain (pre-FFT) and frequency-domain (post-FFT) IA decoding (shown in Figs.~\ref{fig:EVMdegradation_vs_decoderLength}, \ref{fig:EVM_vs_Npilots}, and \ref{fig:EVM_vs_feedbackTime}) translates into a noticeably higher gap in terms of sum-rate, as shown in Fig.~\ref{fig:BERvsSR_PreVsPostFFT}. Note also that more sophisticated algorithms may help reduce the gap between pre-FFT and post-FFT, but this analysis is out of the scope of the paper and we leave it as future work.

\subsection{Comparison of the Adopted Schemes}
\label{sec:comparison}

\begin{figure}[!t]
\centering
\includegraphics[width=\myfigurewidth]{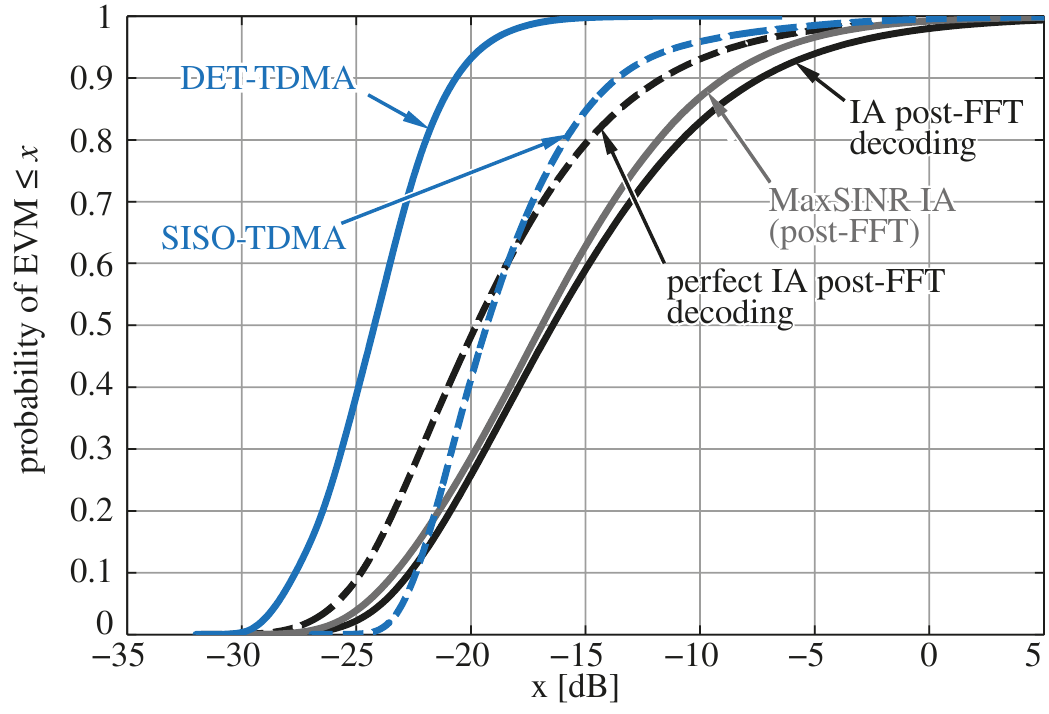}
\caption{{Estimated CDF of the received signal constellation EVM for all the adopted transmission schemes {($M=30$, $L=30$)}.}
}\label{fig:cdf_EVM}
\end{figure}

In this subsection we compare the performance of the five adopted schemes using two different metrics. {With respect to IA, in this subsection we only consider post-FFT (frequency-domain) IA decoding.} First, we show in Fig.~\ref{fig:cdf_EVM} the CDF of the received signal constellation EVM. As expected, DET-TDMA provides the lowest EVM and guarantees an EVM better than $-15$\,dB for all channel realizations, whereas IA ensures the same signal quality in 60\% of the realizations. On the other hand, Fig.~\ref{fig:cdf_EVM} also shows a noticeable degradation of IA with respect to Perfect IA, where the latter is able to achieve the same EVM value of $-15$\,dB in a 20\% more of channel realizations. This effect was already observed in Fig.~\ref{fig:cdf_EVM_preFFT_vs_postFFT} and is due not only to channel estimation errors, which avoid the interference to be perfectly nulled out, but also to transmitter noise and synchronization issues, as already explained in Sections~\ref{sec:channel_realizations}~and~\ref{sec:comparison_preFFT_postFFT}. Alternatively, the MaxSINR scheme provides little EVM improvement over IA, increasing the percentage in only 4\% at $-15$\,dB. This suggests that the operating SNRs are sufficiently high for IA to achieve good performance, and therefore MaxSINR algorithm converges to the zero-forcing IA solution in most subcarriers. However, when there exists high collinearity between the signal and the interference subspaces, MaxSINR enhances the desired channel, thus providing an improvement in the average EVM performance. Finally, it is worth mentioning that the quality of the equivalent channels after applying the IA precoders and decoders, which is represented by the EVM performance of Perfect IA, is more spread than that of SISO channels. This is a reasonable result since IA precoders and decoders are independent of the desired links, hence yielding collinearity as well as orthogonality between the signal and the interference subspaces with the same probability.

\begin{figure}[!t]
\centering
\includegraphics[width=\myfigurewidth]{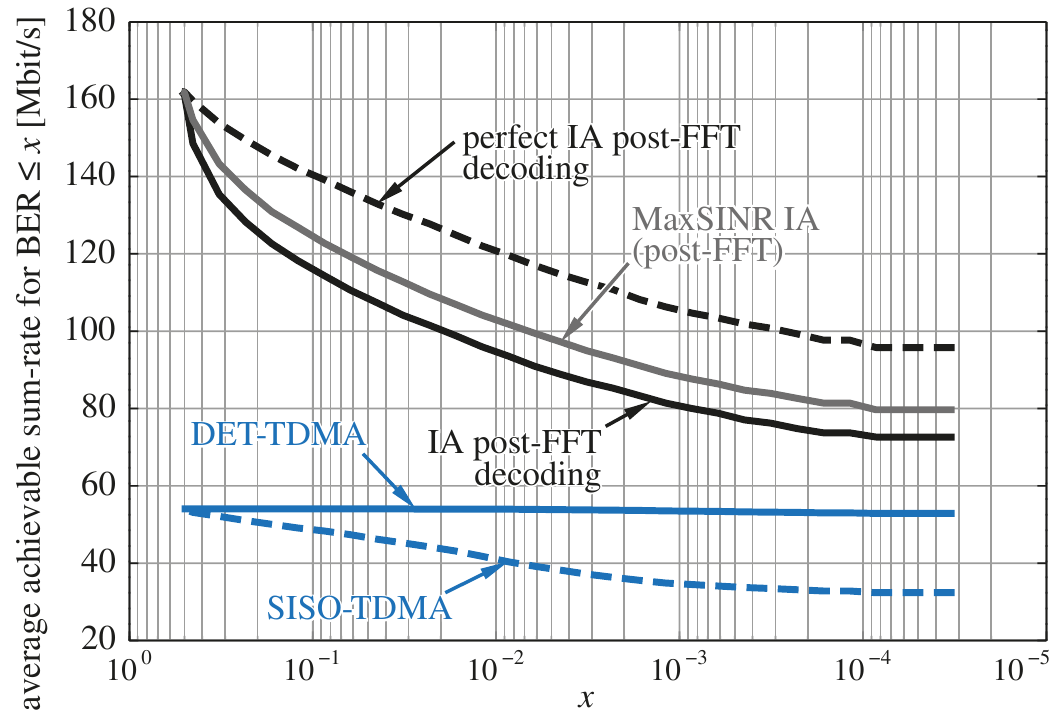}
\caption{{Average achievable sum-rate that guarantees a given BER, assuming optimal rate adaptation {($M=30$, $L=30$)}.}
}\label{fig:BERvsSR}
\end{figure}

Finally, Fig.~\ref{fig:BERvsSR} shows the BER results for the five adopted schemes. We observe that IA schemes achieve higher throughput than TDMA schemes for all BER requirements. For instance, IA provides an average rate of 73\,Mbit/s with a maximum BER of $10^{-4}$, whereas SISO and DET achieve 32 and 53\,Mbit/s, respectively. On the other hand, although MaxSINR does not provide a significant improvement in terms of EVM (see Fig.~\ref{fig:cdf_EVM}) it does provide substantially higher data rates than IA. More specifically, it achieves 7\,Mbit/s more than IA at the same operating point of $\text{BER}\leq 10^{-4}$. This is due to the fact that the channel encoding is sensitive to changes in the received EVM and thus a small improvement in the signal quality may yield a significant BER decrease, hence showing the importance of enhancing the signal quality when collinearity between the signal and the interference subspaces occurs. Following these lines, we also observe that Perfect IA provides a large throughput improvement over IA, which evidences once again the significant impact of practical impairments such as channel estimation errors, transmitter noise and imperfect timing. Such impairments, along with collinearity issues, significantly limit the performance of IA schemes (specially as the number of users increases) and should be considered in future theoretical IA designs.

\section{Conclusions}\label{sec:conclusions}
In this paper we have presented an experimental performance evaluation of spatial IA in the 3-user MIMO-OFDM interference channel and considering an static indoor wireless local area network scenario. We have carefully analyzed the main practical impairments that may degrade the end-to-end performance: imperfect CSI, frame detection in asynchronous scenarios and dirty RF effects. To this end, we have deployed a suitable experimental setup made up of three MIMO transmitters and receivers, and measured received constellation EVM and BER for a set of indoor channels following the conventional frame structure and synchronization strategies of the IEEE 802.11a WLAN standard. We have firstly pointed out that time-domain IA decoding must be applied in totally asynchronous scenarios to cancel out the interference before time synchronization, and we have proposed a simple design for such decoders. Our results indicate that the EVM degradation due to time-domain IA decoding is less than 1\,dB when choosing an appropriate decoder length. Secondly, an analysis of imperfect CSI has been carried out and we have observed that the received EVM is dominated by transmitter noise (dirty RF) when the channel estimates are sufficiently accurate, which significantly limits the end-to-end performance of IA. The performance of IA has also been compared with that of different TDMA schemes, and we have shown that IA may achieve a significantly higher throughput for a given BER requirement under real settings. Finally, this work highlights the relevance of experiments where signals are actually transmitted over the air and all practical impairments are taken into account. This experimental research is not only useful to evaluate theoretical results in real-world scenarios but also to uncover new research lines.

\section*{Acknowledgements}
This work has been supported by Xunta de Galicia, MINECO of Spain, and by FEDER funds of the E.U. under Grant 2012/287, Grant TEC2013-47141-C4-R (RACHEL project), Grant CSD2008-00010 (COMONSENS project), and FPU Grants AP2010-2189 and AP2009-1105.


\bibliographystyle{ieeetr} 
\bibliography{Measurements,InterferenceAlignment}      

\end{document}